\documentstyle[eepic,epsf,11pt]{article}
\def\maru{\hbox{$\hskip0.3em\raisebox{-.0ex}{$~$}
 \raisebox{1.1ex}{\hskip-0.4em$\scriptscriptstyle{\circ}$\hskip0.1em}$}}
\def\gmaru{\hbox{$\hskip0.3em\raisebox{-.0ex}{$g$}
 \raisebox{1.1ex}{\hskip-0.4em$\scriptscriptstyle{\circ}$\hskip0.1em}$}}
\def\etamaru{\hbox{$\hskip0.3em\raisebox{-.0ex}{$\eta$}
 \raisebox{1.1ex}{\hskip-0.4em$\scriptscriptstyle{\circ}$\hskip0.1em}$}}
\def\emaru{\hbox{$\hskip0.3em\raisebox{-.0ex}{$e$}
 \raisebox{1.1ex}{\hskip-0.4em$\scriptscriptstyle{\circ}$\hskip0.1em}$}}
\def\Rmaru{\hbox{$\hskip0.3em\raisebox{-.0ex}{$R$}
 \raisebox{1.8ex}{\hskip-0.4em$\scriptscriptstyle{\circ}$\hskip0.1em}$}}
\def\Fmaru{\hbox{$\hskip0.3em\raisebox{-.0ex}{$F$}
 \raisebox{1.8ex}{\hskip-0.4em$\scriptscriptstyle{\circ}$\hskip0.1em}$}}

\def\bigBox{\raisebox{-0.1em}{\large\boldmath $\Box$}}

\newcommand{\ts}[1]{\textstyle{#1}}
\newcommand{\jb}{{\bar j}}

\newcommand{\NN}{\nonumber}
\newcommand{\BE}{\begin{equation}}
\newcommand{\EE}{\end{equation}}
\newcommand{\BEA}{\begin{eqnarray}}
\newcommand{\EEA}{\end{eqnarray}}
\newcommand{\BEAN}{\begin{eqnarray*}}
\newcommand{\EEAN}{\end{eqnarray*}}
\newcommand{\rot}{{\rm rot}}
\newcommand{\ANN}[3]{Ann. Phys. (NY) {\bf #1} {(#2)} {#3}}
\newcommand{\ATM}[3]{Adv. Theor. Math. Phys. {\bf #1} {(#2)} {#3}}
\newcommand{\CQG}[3]{Class. Quant. Grav. {\bf #1} {(#2)} {#3}}
\newcommand{\CMP}[3]{Comm. Math. Phys. {\bf #1} {(#2)} {#3}}
\newcommand{\JHEP}[3]{JHEP {\bf #1} {(#2)} {#3}}
\newcommand{\NP }[3]{Nucl. Phys. {\bf #1} {(#2)} {#3}}
\newcommand{\PL }[3]{Phys. Lett. {\bf #1} {(#2)} {#3}}
\newcommand{\PR }[3]{Phys. Rev. {\bf #1} {(#2)} {#3}}
\newcommand{\PRL}[3]{Phys. Rev. Lett. {\bf #1} {(#2)} {#3}}
\begin{document}
\baselineskip=18pt
%
\begin{titlepage}
\begin{center}
\vspace*{0.5cm}
{\Large{\bf $D=5$ Simple Supergravity on $AdS_{2}\times S^{3}$}}
\vskip 1.0cm
{\large Akira Fujii\footnote{\tt fujii@tanashi.kek.jp}
 and Ryuji Kemmoku\footnote{\tt kemmoku@tanashi.kek.jp} 
}
\vskip 0.8cm
{\small {\sl Institute of Particle and Nuclear Studies\\
High Energy Accelerator Research Organization (KEK)\\
Tanashi, Tokyo 188-8501, Japan}}
\vskip 0.2cm
\end{center}
\begin{abstract}
The Kaluza-Klein spectrum of $D=5$ simple supergravity compactified on 
$S^3$ is studied. A classical background solution which preserves 
maximal supersymmetry is fulfilled by the geometry of
$AdS_2\times S^3$. The physical spectrum 
of the fluctuations is classified according to 
$SU(1,1|2)\times SU(2)$ symmetry, which has a very similar structure to 
that in the case of compactification on $AdS_{3}\times S^{2}$.
\end{abstract}

\end{titlepage}
\setcounter{footnote}{0}
%
%
\noindent{\bf 1.}\,{\bf Introduction}\quad 
Maldacena's conjecture \cite{Maldacena} is one of 
the most attractive subjects  
among recent progress in the non-perturbative string/M-theory.  
In this paper, we consider $D=5$ simple supergravity (SUGRA) 
with 0-brane solution 
whose near-horizon geometry is the direct product of two-dimensional 
anti-de-Sitter space and three-sphere ($AdS_{2}\times S^{3}$), and 
investigate its Kaluza-Klein spectrum. 
Assuming that the $AdS_{2}/CFT_{1}$ correspondence holds 
\cite{Claus-etal}-\cite{MaldacenaMichelsonStrominger}, we show 
that the spectrum 
possesses 
$SU(1,1|2)\times SU(2)$ symmetry. It must be remarked that 
Gauntlett et.al.\cite{Gauntlett-etal} 
has recently evinced the super-Poincar{\'e} 
group of $AdS_{2}\times S^{3}$ SUGRA to be $SU(1,1|2)\times SU(2)$. 
Our results provide each Kaluza-Klein mode with the precise assignment 
to the certain representation of $SU(1,1|2)\times SU(2)$. 

On the other hand, $D=5$ simple SUGRA possesses a very similar 
structure to that of $D=11$ SUGRA \cite{Cremmer}-\cite{MizoguchiOhta}. 
In particular, the former theory allows 
solitonic string (1-brane) and particle (0-brane) solutions that 
respectively correspond to M5- and M2-brane solutions in the latter
\cite{GibbonsTownsend}. 
While the near-horizon geometry of the solitonic 1-brane in $D=5$ simple 
SUGRA is 
$AdS_{3}\times S^{2}$, that of the solitonic 0-brane is 
$AdS_{2}\times S^{3}$.
In the previous work \cite{FujiiKemmokuMizoguchi}, 
the Kaluza-Klein spectrum 
in this $AdS_{3}\times S^{2}$ compactification is studied and its 
$SU(1,1|2)_{\rm R}\times SU(1,1)_{\rm L}$ symmetry, 
which can be regarded 
as the finite-dimensional subalgebra of chiral $N=(4,0)$ 
superconformal algebra,
is found.
Interestingly enough, 
two different compactifications are endowed with quite similar 
symmetries, chiral $SU(1,1|2)_{\rm R}\times SU(1,1)_{\rm L}$ and single 
$SU(1,1|2)\times SU(2)$. 
It originates in the magnetic/electric duality between 
$AdS_{3}\times S^{2}$ and $AdS_{2}\times S^{3}$ simple SUGRA theories. \\
~\\
%
%
{\bf 2.}\,{\bf Field equations}\quad 
In terms of the metric $g_{MN}$, $U(1)$ gauge field $A_{M}$, and 
the spin-{3/2} field $\psi_{M}$, $D=5$ simple SUGRA is 
defined by the Lagrangian
\begin{eqnarray}
{\cal L}&=&e_{5}\left[\,-\frac14 R-{1\over 4}F^{MN}F_{MN}\right.\nonumber\\
&&-\frac i2\left(\overline{\psi}_M\widetilde{\Gamma}^{MNP}
         D_N\left(\frac{3\omega-\widehat{\omega}}2\right)\psi_P
         +\overline{\psi}_P\stackrel{\leftarrow}{D}_N
         \left(\frac{3\omega-\widehat{\omega}}2\right)
         \widetilde{\Gamma}^{MNP}\psi_M\right)\nonumber\\
&&-\frac1{6\sqrt{3}}e_5^{-1}\epsilon^{MNPQR}F_{MN}F_{PQ}A_R\nonumber\\
&&\left.-\frac{\sqrt{3}i}8\psi_M(\widetilde{\Gamma}^{MNPQ}
+2g^{M[P}g^{Q]N})\psi_N(F_{PQ}+\widehat{F}_{PQ})\right],
\label{eq:ssugra}
\end{eqnarray}
where the notations used here follow those in \cite{ChamseddineNicolai} and  
the signature is $(+----)$. 
Capital Roman letters run from $0$ to $4$. 
In the absence of the $\psi_{M}$ field, Einstein-Maxwell's 
equation is shown to be
\begin{eqnarray}
&& R_{MN}-\frac12 g_{MN}R = -\left(2F_{MP}F_N^{~~P}-\frac12 g_{MN}F^2\right),
\nonumber\\
&&F^{MN}_{~~~~;M}=\frac1{2\sqrt{3}}e_{5}^{-1}\epsilon^{NPQRS}F_{PQ}F_{RS}.
\label{einsteinmaxwell}
\end{eqnarray}
The above equation of motion enjoys Freund-Rubin-like solutions 
\cite{FreundRubin} that preserve the maximal number of the supersymmetries 
after three-dimensional compactification. 
One of these such solutions over which we will be 
considering the Kaluza-Klein spectrum possesses the geometry of 
$AdS_{2}\times S^{3}$, namely,
\begin{eqnarray}
\Rmaru_{\mu\nu\rho\sigma}&=&\ts{4\over 3}f^2
\left[ \gmaru_{\mu\rho}\gmaru_{\nu\sigma}
                              -\gmaru_{\mu\sigma}\gmaru_{\nu\rho}\right] , 
\nonumber\\
\Rmaru_{mnpq}&=&-\ts{1\over 3}f^2
\left[ \gmaru_{mp}\gmaru_{nq}-\gmaru_{mq}\gmaru_{np}\right] ,
 \\
\gmaru_{\mu m}=\Fmaru_{\mu m}=0
&,&\Fmaru_{\mu\nu}=f\emaru_2\epsilon_{\mu\nu},\,\,\Fmaru_{mn}=0,
\nonumber
\label{AdS3xS2}
\end{eqnarray}
where 
Greek letters are used for $0$ or $1$, small Roman letters 
for $2$,$3$ or $4$, and $\maru$ means the background.
Let us set the free parameter $f$ to be $\sqrt{3}$.\\
~\\
%
%
{\bf 3.}\,{\bf Kaluza-Klein spectrum: bosonic modes}\quad 
The spectrum of small variations around the above $AdS_{2}\times S^{3}$ 
background is derived from  Einstein-Maxwell's equation 
(\ref{einsteinmaxwell}). We docket the small variations of the 
bosonic fields, $g_{MN}$ and $A_{M}$, like
\begin{equation}
\delta g^{MN}=h^{MN},\qquad \delta A_{M}=a_{M}.
\end{equation}
After fixing the gauge and diffeomorphism degrees of freedom in the way 
that 
\begin{equation}
a_m^{~~;m}=0, \quad 
h^{\mu m}_{~~~;\mu}=0, \quad 
h^{\mu m}_{~~~;m}=0,\quad h^{m}_{~~~m}=0,
\end{equation}
the equation of motion for the bosonic fields reads
\begin{eqnarray}
&&-\frac12(h^{\lambda}_{~\mu;\nu;\lambda}+h^{\lambda}_{~\nu;\mu;\lambda}
         -h^{\lambda}_{~~\lambda;\mu;\nu}-h_{\mu\nu~~~;M}^{~~~;M})
\nonumber\\
&&\hspace{1.5cm}+4h_{\mu\nu}-5\gmaru_{\mu\nu}h^{\lambda}_{~\lambda}
+\frac12\gmaru_{\mu\nu}(h^{\mu\nu}_{~~~;\mu;\nu}+h^{mn}_{~~~;m;n}
-h^{\lambda~;M}_{~\lambda~~;M})\nonumber\\
&&\hspace{2.5cm}+4\sqrt 3\{ \emaru_2\epsilon^{~\lambda}_{\mu}
a_{[\lambda;\nu]}+\emaru_2\epsilon^{~\lambda}_{\nu}a_{[\lambda;\mu]}\}
-2\sqrt 3\gmaru_{\mu\nu}\emaru_2\epsilon^{\lambda\sigma}
a_{\sigma;\lambda}=0,
\label{E1}
\\
&&-\frac12(h^{\lambda}_{~\mu;m;\lambda}+h^r_{~m;\mu;r}
         -h^{\lambda}_{~~\lambda;\mu;m}-h_{\mu m~~;M}^{~~~;M})
-3 h_{\mu m}+4\sqrt{3}\emaru_2\epsilon_{\mu}^{~\lambda}a_{[\lambda;m]}
=0,
\label{E2}
\\
&&-\frac12(h^{r}_{~m;n;r}+h^r_{~n;m;r}
         -h^{\lambda}_{~~\lambda;m;n}-h_{mn~~;M}^{~~~;M})\nonumber\\
&&\hspace{1.5cm}-2h_{mn}+\gmaru_{mn}h^{\lambda}_{~\lambda}
+\frac12\gmaru_{mn}(h^{\mu\nu}_{~~;\mu;\nu}+h^{mn}_{~~;m;n}
-h^{\lambda~;M}_{~\lambda~;M})\nonumber\\
&&\hspace{2.5cm}-2{\sqrt{3}}\gmaru_{mn}\emaru_2\epsilon^{\lambda\sigma}
a_{\sigma;\lambda}=0, \\
\label{E3}
&&a^{\mu;M}_{~~~~;M}-4a^{\mu}-\frac{\sqrt{3}}2 
\emaru_2\epsilon^{\mu\sigma}h^{\lambda}_{~\lambda;\sigma}=0,
\label{M1}
\\
&&a^{m;M}_{~~~~;M}+2a^m+4\emaru_3\epsilon^{mpq}a_{q;p}
+\sqrt 3\emaru_2\epsilon^{\lambda\sigma}h^m_{~~\sigma;\lambda}=0.
\label{M2}
\end{eqnarray}
To diagonalize the mass matrices derived from the above equation,  
we adopt the spherical harmonics on $S^{3}$ 
with the rank-0, 1 and 2. 
To begin with, let us summarize the properties of three-dimensional 
spherical harmonics. 
Because the isometry group of $S^{3}$ is $SO(4)$, 
eigenfunctions of the three-dimensional Laplacian 
$\Delta=\nabla^{m}\nabla_{m}$, and the three-dimensional spherical 
harmonics, simultaneously belong to a certain representation of $SO(4)$. 
On the other hand, $SO(4)$ is decomposed into a direct 
product of two $SU(2)$'s, namely $SU(2)\times\overline{SU(2)}$. 
Therefore, we are able to use spins of two $SU(2)$'s  
to classify the Kaluza-Klein spectrum. Let 
${\vec J}^{2}$ (${\vec{\bar J}}^{2}$) denote 
the Casimir operator of $SU(2)$ ($\overline{SU(2)}$)
with eigenvalues 
$j(j+1)$ (${\bar j}({\bar j}+1)$) for $j,{\bar j}=0,1/2,1,3/2,\cdots$. 
$\Phi^{(j)}$ (${\bar \Phi}^{({\bar j})}$) is the
 representation of $SU(2)$ ($\overline{SU(2)}$) with spin 
$j$ (${\bar j}$). In general, $j$ and $\jb$ can take different 
half-integer values. The difference $|j-\jb|$ is called the rank of the 
spherical harmonic. This rank corresponds to that of the tensor structure 
not on $AdS_{2}$ but on $S^{3}$. Spherical harmonics with rank-0,1, and 2 
are explicitly constructed as follows. 
\begin{enumerate}
\item rank-0 harmonics\\
First, for scalar functions on $S^{3}$, one can confirm 
the equality among ${\vec J}^{2}$, ${\vec{\bar J}}^{2}$, and $\Delta$
\begin{equation}
-\Delta=4{\vec J}^{2}=4{\vec{\bar J}}^{2}. 
\end{equation}
It is shown that a product 
$Y^{(k)}=\Phi^{(k/2)}{\bar \Phi}^{(k/2)}$ behaves as  
a scalar on $S^{3}$ and is an eigenfunction of $\Delta$. 
In fact, its eigenvalue for $\Delta$ is given 
by $-k(k+2)=-4j(j+1)=-4\jb (\jb +1)$, 
where $k=2j=2\jb =0,1,2,\cdots$.  
\item rank-1 harmonics\\ 
In a similar way to the case for rank-0 harmonics, 
one can prove that   
$Y_{m}^{(k,\pm)}=\Phi^{({k\pm 1\over 2})}{\bar \Phi}^{({k\mp 1\over 2})}$ 
are vector 
functions on $S^{3}$ and, simultaneously, eigenfunctions 
of the Laplacian with the 
eigenvalue $-\Delta=k(k+2)-1$ for $k=1,2,3,\cdots$.
Introducing the 
rotational derivative for vector fields 
$(\rot\,v)^{m}=\emaru_{3}\epsilon^{mnp}v_{p;n}$, 
we should remark that 
\begin{equation}
\rot\,Y_m^{(k,\pm)}=\pm(k+1) Y_m^{(k,\pm)}.
\end{equation} 
\item rank-2 harmonics\\  
$Y_{mn}^{(k,\pm)}=\Phi^{({k\pm 2\over 2})}{\bar \Phi}^{({k\mp 2\over 2})}$ 
are rank-2 tensors   
on $S^{3}$ and are eigenfunctions of $\Delta$ with the 
eigenvalue $-\Delta=k(k+2)-2$ for $k=2,3,4,\cdots$.
\end{enumerate}

By expanding the bosonic fields $a_{M}$ and $h_{MN}$ in terms of the above 
spherical harmonics, the field equation (\ref{E1})-(\ref{M2}) is 
separated into three sets with rank-0, 1 and 2.
The $SO(4)$-charge of each bosonic mode is inherited from the used spherical 
harmonic with a certain rank. The results are summarized in the following.
\begin{enumerate}
\item rank-0\\ 
The eigenvalues of the two-dimensional 
d'Alembertian $\bigBox$ on $AdS_{2}$ for
$h^{mn}_{~~~;m;n}$, $h^{\mu\nu}_{~~;\mu;\nu}$, 
$\etamaru^{\lambda\sigma}a_{\sigma;\lambda}$ and $h^{\lambda}_{\ \lambda}$
are 
\begin{equation}
\lambda^{2}=-\bigBox=k^{2}-2k \quad {\rm for}\quad k=2,3,4,\cdots 
\label{eq:tower1} 
\end{equation}
and 
\begin{equation}
\lambda^{2}=k^{2}+6k+8 \quad {\rm for}\quad k=0,1,2,\cdots. 
\label{eq:tower2}
\end{equation} 
The reason why 
the tower (\ref{eq:tower1}) begins from $k=2$ is because 
at $k=0$ the mass-matrix is degenerated, and at $k=1$ the eigenvalue 
corresponding to this tower is inconsistent with the tensor structure of 
$h^{mn}$ which is the origin of $h^{mn}_{~~~;m;n}$. 
The $SO(4)$-charges corresponding to both of the above two branches prove 
identically 
\begin{equation}
(j,\jb)=(\ts{\frac k2}, \ts{\frac k2}).
\end{equation} 
\item rank-1\\ 
Let us consider the mass-matrices on 
$\etamaru^{\lambda\sigma}h_{m\lambda;\sigma}$ and $a_m$. 
One can immediately see 
that the mass matrices take different forms for 
$Y_{m}^{(k,+)}$ or $Y_{m}^{(k,-)}$. The eigenvalues for 
$Y_{m}^{(k,+)}$ are
\begin{equation}
\lambda^{2}=k^{2}+8k+15\quad {\rm or}\quad k^2-1
\end{equation}
with the same $SO(4)$ charge 
\begin{equation}
(j,\jb)=(\ts{k+1\over 2},\ts{k-1\over 2})
\end{equation}
for $k=1,2,3,\cdots$. The eigenvalues for $Y_{m}^{(k,-)}$ are
\begin{equation}
\lambda^{2}=k^{2}-4k+3 \quad {\rm for}\quad k=2,3,4\cdots
\label{eq:tower13}
\end{equation}
and
\begin{equation}
\lambda^{2}=k^{2}+4k+3 \quad {\rm for}\quad k=1,2,3\cdots
\end{equation}
with the $SO(4)$-charges
\begin{equation}
(j,\jb)=(\ts{k-1\over 2},\ts{k+1\over 2}).
\end{equation}
We should remark that 
the eigenvalue (\ref{eq:tower13}) for $k=1$ is missing. 
At $k=1$, this mode is massless and degenerates into that of
$k^2-1\ (k=1)$. 
However, the dynamical degree of freedom for massless scalars 
in the present case
is merely one, and the three-dimensional chirality such that $j-\bar{j}<0$
is not consistent with the two-dimensional parity of the scalar fields.
Therefore, we drop that mode.
\item rank-2\\
On the rank-2 field $h_{mn}$, the d'Alembertian has the eigenvalue 
\begin{equation}
\lambda^{2}=k^{2}+2k \quad {\rm for}\quad k=2,3,4\cdots,
\end{equation}
for both  
\begin{equation}
(j,\jb)=(\ts{k\pm 2\over 2},\ts{k\mp 2\over 2}). 
\end{equation}
\end{enumerate}
~\\
%
%
{\bf 4.}\,{\bf Kaluza-Klein spectrum: fermionic modes}\quad 
The fermionic fluctuation around the $AdS_{2}\times S^{3}$ background, 
$\psi_{M}=0$, is labeled simply by $\psi_{M}$ itself. 
Its linearized equation  
of motion is drawn from the Lagrangian (\ref{eq:ssugra}):
\begin{eqnarray}
i\Gamma^{MNP}\psi_{P;N}
-\frac{\sqrt{3}}{4}i(\Gamma^{MNPQ}
                     +2\gmaru^{M[P}\gmaru^{Q]N})\psi_N \Fmaru_{PQ}=0.
\label{5dRSeq}
\end{eqnarray}
To analyze the Kaluza-Klein spectrum from (\ref{5dRSeq}), 
we utilize rank-$1/2$ spherical harmonic $\xi^m$ on $S^{3}$ 
which is an eigenspinor of the equation 
\begin{equation}
\Gamma^{n}\nabla_{n}\xi^m +\Gamma^{mn}\xi_n= \kappa\,\xi^m. 
\label{eq:rank1/2}
\end{equation}
Once such rank-$1/2$ spherical harmonic $\xi^{m}$ and eigenvalue 
$\kappa$ are found, the field equation (\ref{5dRSeq}) is reduced 
to the following two-dimensional Dirac equation for $\psi^{m}$, 
\begin{equation}
\gamma^5\gamma^{\mu}\psi^m_{\ ;\mu}+\left(\kappa-{1\over 2}\right)\psi^m=0.
\label{eq:2dDirac}
\end{equation}
Therefore, the fermionic spectrum is obtained by solving the eigenvalue 
problem (\ref{eq:rank1/2}). 

The rank-$1/2$ spherical harmonic $\xi^{m}$ can be expanded by the 
product of rank-1 harmonics and Killing spinors on $S^{3}$, {\it i.e.} 
the seperation of the total angular momentum into the orbital and 
the internal ones. By proceeding this expansion with the help of 
the gauge fixing $\Gamma_M\psi^M=0$, we see that the eigenvalue 
$\kappa$ can take one of the values
\begin{eqnarray}
\kappa=-k-\ts{5\over 2},\quad 
(j,\jb)=(\ts{k+2\over 2},\ts{k-1\over 2})
&&{\rm for}\quad k=1,2,3\cdots,\label{eq:k1}\\
\kappa=k-\ts{1\over 2},\quad
(j,\jb)=(\ts{k-2\over 2},\ts{k+1\over 2})
&&{\rm for}\quad k=2,3,4\cdots,\label{eq:k2}\\
\kappa=-k-\ts{1\over 2},\quad
(j,\jb)=(\ts{k\over 2},\ts{k-1\over 2})
&&{\rm for}\quad k=1,2,3\cdots,\label{eq:k3}\\
\kappa= k+\ts{3\over 2},\quad
(j,\jb)=(\ts{k\over 2},\ts{k+1\over 2})
&&{\rm for}\quad k=1,2,3\cdots.\label{eq:k4}
\end{eqnarray}
Several remarks are in order. First, 
for the two towers (\ref{eq:k1}) and (\ref{eq:k2}), the 
{\it gamma-traceless} condition,
$\Gamma_m\xi^m=0$, which means $|j-\jb|=3/2$, is satisfied  
while it is not satisfied for the other two. 
Secondly, the eigenstate for (\ref{eq:k2}) at $k=1$ is absent 
because one can 
show that for that tower, $\xi^m=0$ at $k=1$ by using its explicit form.

The way to assign the $SO(4)$-charge to each mode indicated in 
(\ref{eq:k1})-(\ref{eq:k4}) may need 
some explanation. 
To see this, one should utilize the above mentioned fact that 
$\xi^{m}$ can be expanded by the product of $Y^{(k,\pm)}_{m}$  
with $SO(4)$-charge $(k/2\pm 1/2,k/2\mp 1/2)$ 
and Killing spinors with $(1/2,0)$. Let us note here that 
the Killing spinors with $(1/2,0)$ possess the negative chirality for  
${\rm R}^{4}$ in which $S^{3}$ is embedded.
In summary, 
the rank-$1/2$ spherical harmonic $\xi^{m}$ has 
one of the $SO(4)$ charges in the following decomposition.  
\begin{eqnarray}
(\ts{k+1\over 2},\ts{k-1\over 2})\otimes 
(\ts{1\over 2},0)&\oplus&
(\ts{k-1\over 2},\ts{k+1\over 2})\otimes 
(\ts{1\over 2},0)\NN \\
&=&(\ts{\frac{k+2}2},\ts{\frac{k-1}2})\oplus
(\ts{\frac{k-2}2},\ts{\frac{k+1}2)}\oplus
(\ts{\frac{k}2},\ts{\frac{k-1}2})\oplus
(\ts{\frac{k}2},\ts{\frac{k+1}2}),
\label{direct-sum}
\end{eqnarray}
where the representation as $(\ts{\frac{k+1}2},\ts{\frac{k-2}2})$
starts from $k=2$ while the others from $k=1$. The irreducible decomposition 
(\ref{direct-sum}) gives the charges in (\ref{eq:k1})-(\ref{eq:k4}).\\
~\\
%
%
{\bf 5.}\,{\bf Symmetry of the spectrum}\quad 
Let us classify the acquired Kaluza-Klein spectrum by a certain 
symmetry group. By considering the super-Poincar{\' e} group consisting 
of the isometry and the supercharges, the symmetry of the modes 
is shown to be a  
superalgebra $SU(1,1|2)\times SU(2)$ \cite{Gauntlett-etal}. 
We should remark 
that the bosonic part of $SU(1,1|2)\times SU(2)$ is 
$SU(1,1)\times SU(2)\times SU(2)\cong SO(1,2)\times SO(4)$, 
which is nothing but the product of the isometry group 
of $AdS_2$ and that of $S^3$. Moreover, let us note that 
$SU(1,1|2)$ is a finite-dimensional 
subalgebra of $N=4$ super-Virasoro algebra.
The oscillator representations \cite{Guenaydin}, 
which is very helpful to visualize the representations,  
of $SU(1,1|2)$ has already appeared in the 
analysis of the Kaluza-Klein spectrum on $AdS_{3}\times S^{3}$ 
 and also on $AdS_{3}\times S^{2}$ \cite{deBoer}-\cite{Larsen}. 

Assuming that this SUGRA is coupled to a one-dimensional 
conformal field theory at the boundary of $AdS_{2}$, we can 
translate each Kaluza-Klein mode  
into a conformal weight ${\bf h}$ by means of 
the formula in \cite{BreitenlohnerFreedman,Witten}
\begin{equation}
{\bf h}=\frac{1+\sqrt{1+\lambda^2}}2.
\end{equation}
Hence we can make Table 1.
\vskip 2ex

\renewcommand{\arraystretch}{1.2}
\begin{center}
\begin{tabular}{|l|c|c|c|c|}
\hline
\hspace*{1.6cm}$\lambda^{2}$ & ${\bf h}$ & $(j,\jb)$ & $j-\jb$&{\tt Fig.}\\
\hline\hline
$k^{2}-2k\ (k\geq 2)$ & $\frac{k}2$ & $(\frac k2,\frac k2)$ & $0$ 
&{\tt 1} \\
$k^{2}+6k+8\ (k\geq 0)$& $\frac{k+4}2$ & $(\frac k2,\frac k2)$ & $0$ 
& {\tt 2} \\
$k^{2}+4k+3\ (k\geq 1)$& $\frac{k+3}2$& $(\frac {k-1}2,\frac {k+1}2)$& $-1$ 
&{\tt 1}  \\
$k^{2}-1\ (k\geq 1)$& $\frac{k+1}2$ & $(\frac {k+1}2,\frac {k-1}2)$ & $1$
&{\tt 2} \\
$k^{2}-4k+3\ (k\geq 2)$& $\frac{k-1}2$& $(\frac {k-1}2,\frac {k+1}2)$&$-1$
&{\tt 3}\\
$k^{2}+8k+15\ (k\geq 1)$& $\frac{k+5}2$& $(\frac {k+1}2,\frac {k-1}2)$ 
& $1$&{\tt 4} \\
$k^{2}+2k\ (k\geq 2)$& $\frac{k+2}2$& $(\frac {k-2}2,\frac {k+2}2)$ & $-2$
&{\tt 3} \\
$k^{2}+2k\ (k\geq 2)$& $\frac{k+2}2$& $(\frac {k+2}2,\frac {k-2}2)$ & $2$
&{\tt 4} \\
\hline
\end{tabular}\\
\vskip 1ex
Table 1:\ Bosonic modes.
\end{center}

\vskip 2ex

On the other hand, we can carry on the same mapping 
for the fermionic modes by using the asymptotic form of $\psi^{m}$ 
determined by (\ref{eq:2dDirac}),
\begin{equation}
{\bf h}=\frac12\left|\kappa-\frac12\right|+\frac12.
\end{equation}
The table for the fermionic modes turns out as Table 2.

\vskip 2ex

\begin{center}
\begin{tabular}{|r|c|c|c|c|}
\hline
$\kappa$\hspace{1.2cm} & ${\bf h}$ & $(j,\jb)$ & $j-\jb$&{\tt Fig.} \\
\hline\hline
$k+\frac32\ (k\geq 1)$& $\frac{k+2}2$&$(\frac {k}2,\frac {k+1}2)$& 
$-\frac12$&{\tt 1}  \\
$-k-\frac12\ (k\geq 1)$& $\frac{k+2}2$&$(\frac {k}2,\frac {k-1}2)$& 
$\frac12$&{\tt 2} \\
$k-\frac12\ (k\geq 2)$ & $\frac{k}2$& $(\frac {k-2}2,\frac {k+1}2)$&
 $-\frac32$&{\tt 3}  \\
$-k-\frac52\ (k\geq 1)$ & $\frac{k+4}2$& $(\frac {k+2}2,\frac {k-1}2)$
 &$\frac32$&{\tt 4} \\
\hline
\end{tabular}\\
\vskip 1ex
Table 2:\ Fermionic modes.
\end{center}

\vskip 2ex
Finally we can fit 
these results into representations of
$SU(1,1|2)\times\overline{SU(2)}$.
In \cite{FujiiKemmokuMizoguchi}, 
we dealt with a chiral $SU(1,1|2)_{R}\times SU(1,1)_{L}$ 
symmetry.  
Upon replacing the $SU(1,1)_{L}$ by $SU(2)$,
we confirm that all Kaluza-Klein modes fall into four 
supermultiplets of $SU(1,1|2)\times SU(2)$ in a similar manner 
to the case of $AdS_3\times S^2$. 
Every spectrum obeying this single 
$SU(1,1|2)\times SU(2)$ symmetry is summarized in 
Figs. 1-4. In those four figures, one of the two $SU(2)$-charges, 
namely $\jb$ is fixed at a certain value. For example, the fields 
with the same $\jb = k/2$ and $j-\jb =0,-1/2$, and $-1$ are gathered in 
Fig.1. In each figure, one can notice a multiplet-shortening very similar 
to that for the chiral primary fields in two-dimensional $N=4$ 
superconformal theories.
The shortest supermultiplet appears as a doubleton in Fig. 3 $(k=1)$.
The massless multiplets exist in Figs. $1,2,3\ (k=2)$.\\
~\\
%
%
{\bf 6.}\,{\bf Discussions}\quad 
 From the explicit calculation of the Kaluza-Klein spectrum 
of the $AdS_{2}\times S^{3}$ simple SUGRA, we have seen that 
the two different compactifications, 
$AdS_{3}\times S^{2}$ and $AdS_{2}\times S^{3}$, are 
characterized by closely related superalgebras, 
$SU(1,1|2)_{\rm R}\times SU(1,1)_{\rm L}$ and 
$SU(1,1|2)\times SU(2)$, respectively, each of which allows short 
representations. The similarity between the spectra of 
$AdS_{3}\times S^{2}$ and $AdS_{2}\times S^{3}$ is not accidental but 
owing to the magnetic/electric duality for the solitonic 
1-brane and the 0-brane in $D=5$ simple SUGRA. Einstein-Maxwell's equation 
(\ref{einsteinmaxwell}) allows two kinds of solitonic objects. 
One is the solitonic 1-brane solution as follows. 
The metric with a solitonic 1-brane is  
\begin{equation}
ds_{5}^{2}=H_1^{-1}(-dt^{2}+dy^{2})
         +H_1^{2}(dr^{2}+r^{2}d\Omega_{2}^{2}),\quad H_1=1+{Q_1\over r},
\end{equation}
where $(t,y)$ is the coordinate of the world-sheet, $r$ is the distance 
from the soliton, and $Q_{1}$ is a constant. 
The non-zero field strength of the gauge field is 
\begin{equation}
F^{ij}=-\sqrt{3}\epsilon^{ijk}H_1^{-4}\partial_{k}H_1,
\end{equation}
where both indices $i$ and $j$ denote transverse directions, 
{\it i.e.} not $t$ or $y$.
>From the form of the field strength, this soliton corresponds to a  
magnetic monopole. It is worthwhile to mention that the near-horizon 
geometry is nothing but $AdS_{3}\times S^{2}$. 
The other one is the solitonic 0-brane solution. The metric and the 
field strength in this configuration are 
\begin{eqnarray}
ds_{5}^{2}&=&-H_0^{-2}dt^{2}
         +H_0(dr^{2}+r^{2}d\Omega_{3}^{2}),\NN\\
F^{\mu\nu}&=&(\delta^{\mu}_{t}\delta^{\nu}_{r}-\delta^{\mu}_{r}
\delta^{\nu}_{t})
\cdot\sqrt{3}{Q_0\over r^{3}H_0},\\
H_0&=&1+{Q_0\over r^{2}},\NN
\end{eqnarray}  
where $t$ parameterizes the world-line, $r$ is the 
distance from the 0-brane, and $Q_{0}$ is the charge of the soltonic 0-brane. 
One can see that the gauge field 
corresponds to that induced by an electric charge and 
that the near-horizon geometry  
turns out to be $AdS_{2}\times S^{3}$. 
Therefore, the similarity of the spectra 
can be regarded as a reminiscent of the five-dimensional magnetic/electric 
duality. 

On the other hand, as is discussed in \cite{Strominger}, 
the $SU(1,1)$ isometry of $AdS_{2}$ can be enlarged to a single 
infinite-dimensional Virasoro symmetry in the boundary similarly 
to that in the case of $AdS_{3}/CFT_{2}$ \cite{BrownHenneaux}. 
This single (super-)Virasoro 
algebra has been explicitly constructed in the study of  
{\it conformal mechanics} that describes the behavior of a test particle 
near the horizon of a four-dimensional extreme 
Reissner-Nordstr{\"o}m black hole \cite{Claus-etal}. 
It would be interesting to explore the relation between 
the conformal mechanics 
and $AdS_{2}\times S^{3}$ simple SUGRA. \\

\noindent{\bf Acknowledgements}\\
The authors would like to thank Shun'ya Mizoguchi for his 
helpful discussions and comments on the manuscript. They also acknowledge  
Hisaye Hosokawa for her reading the manuscript. 
 
%

%
%
\newpage
\vspace*{4cm}
\large{\bf{Figure Captions}}\\

\begin{tabular}{ll}
Fig.1 & Multiplet of boundary fields with 
$j-\bar{j}=0,\ts{-\frac12},-1$. \\
&\\
Fig.2 & Multiplet of boundary fields with 
$j-\bar{j}=1,\ts{\frac12},0$. \\
&\\
Fig.3 & Multiplet of boundary fields with 
$j-\bar{j}=-1,\ts{-\frac32},-2$. \\
&\\
Fig.4 & Multiplet of boundary fields with 
$j-\bar{j}=2,\ts{\frac32},1$. \\
\end{tabular}
%
%
\newpage
\begin{figure}
  \epsfxsize = 12 cm   
  \centerline{\epsfbox{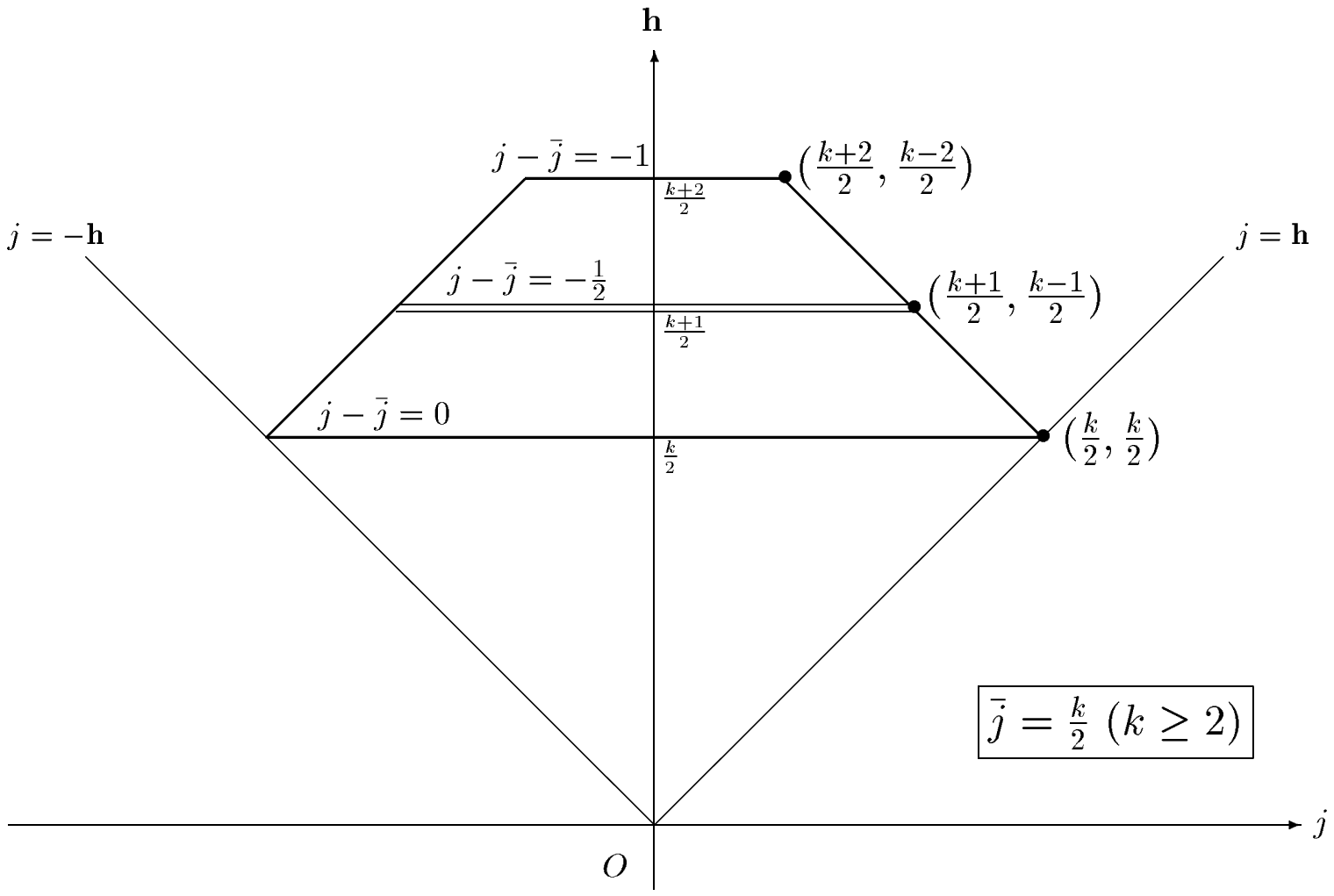}}
  \centerline{~}
  \centerline{\large{\tt Fig.1}}
\end{figure}
\begin{figure}
  \epsfxsize = 12 cm   
  \centerline{\epsfbox{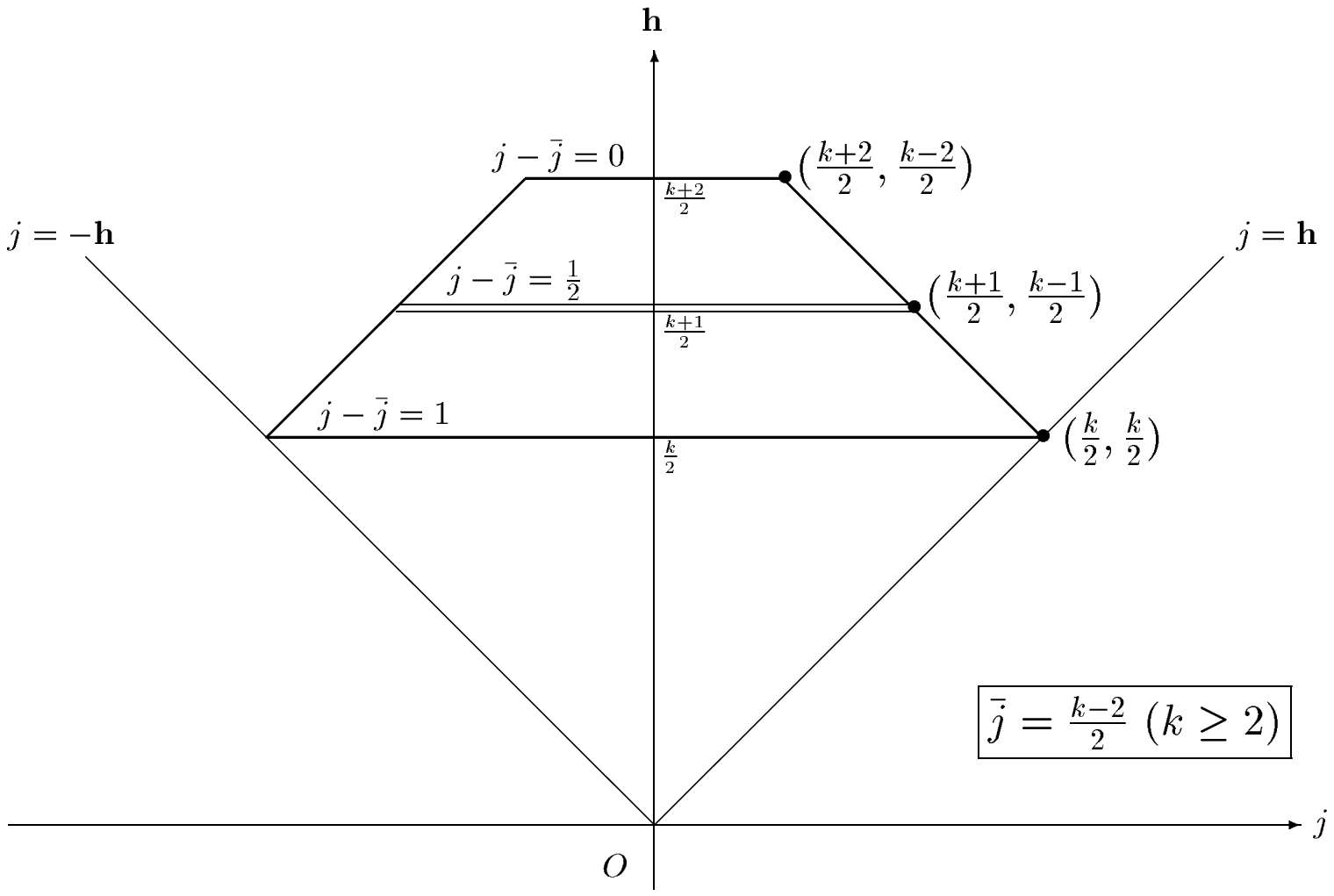}}
  \centerline{~}
  \centerline{\large{\tt Fig.2}}
\end{figure}
\begin{figure}
  \epsfxsize = 12 cm   
  \centerline{\epsfbox{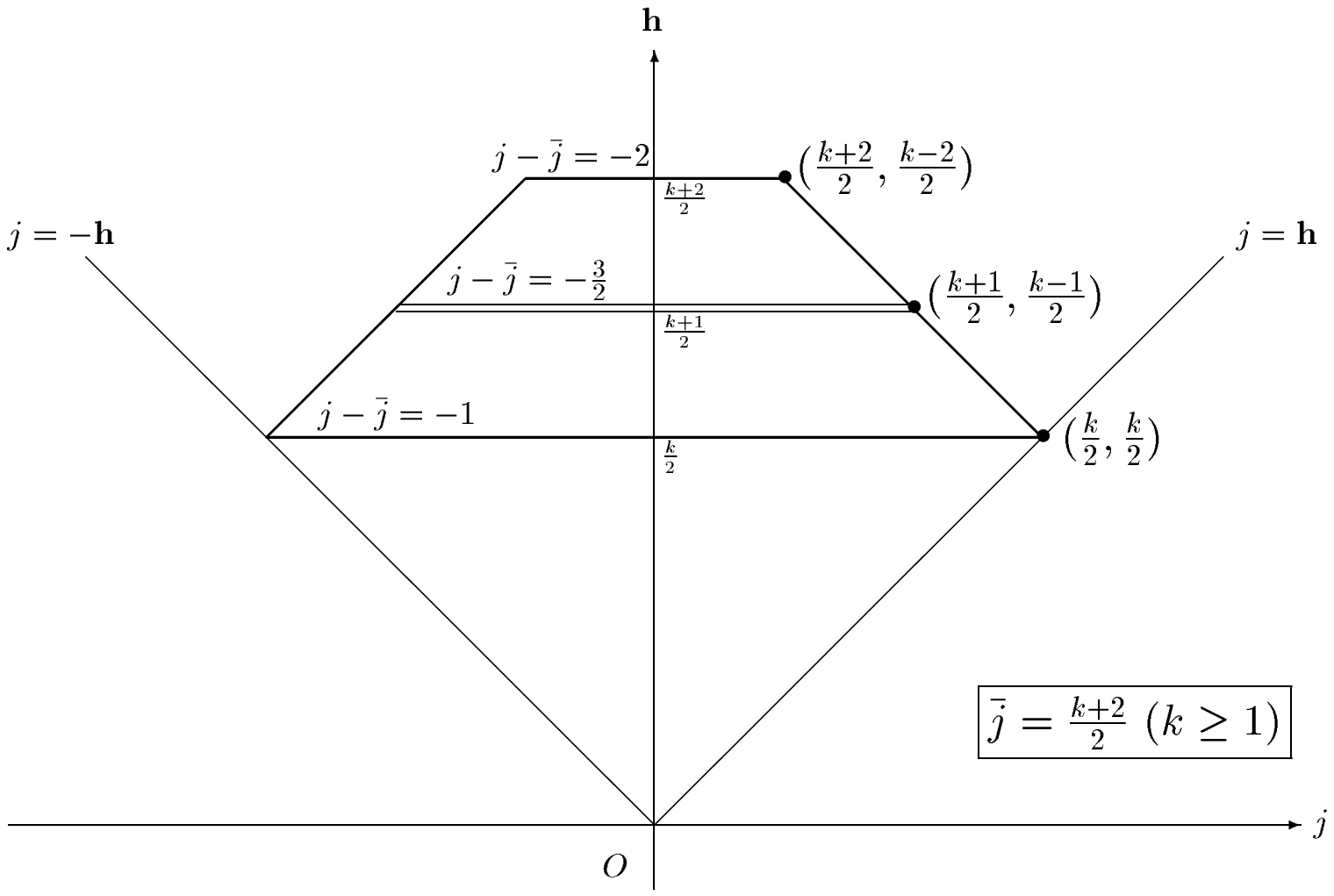}}
  \centerline{~}
  \centerline{\large{\tt Fig.3}}
\end{figure}
\begin{figure}
  \epsfxsize = 12 cm   
  \centerline{\epsfbox{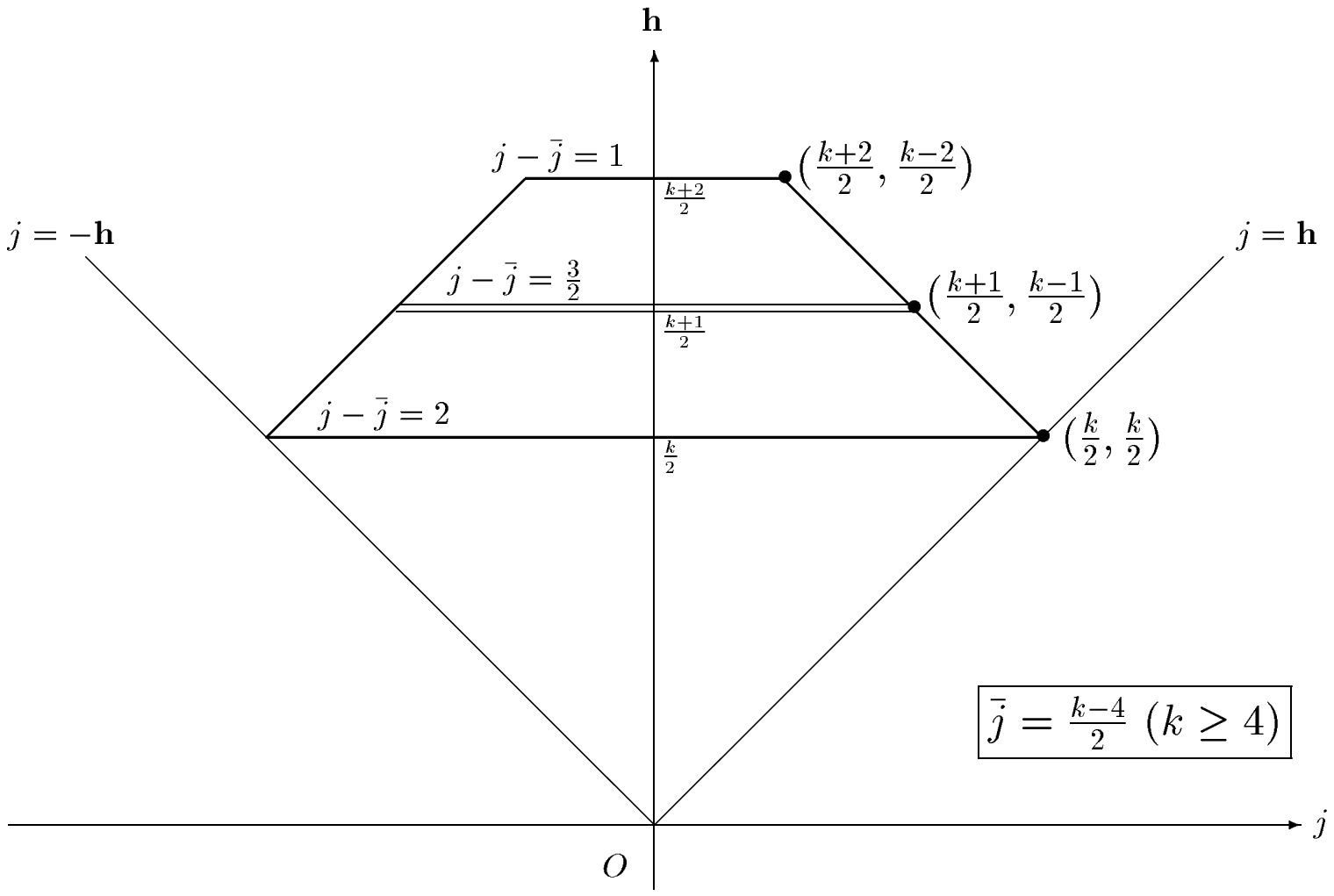}}
  \centerline{~}
  \centerline{\large{\tt Fig.4}}
\end{figure}

\end{document}